\documentclass[preprint,aps,pra,showpacs,floatfix]{revtex4}
\usepackage{graphicx}
\usepackage{nicefrac}
\usepackage{amsmath}
\usepackage{amsfonts}
\usepackage{amssymb}
\usepackage{amsthm}
\usepackage{epsf}
\usepackage{bm}

\usepackage{dcolumn}
\newcolumntype{.}{D{x}{}{-1}}

\newcommand{\rp}{{r^{\prime}}}
\newcommand{\rpp}{{r^{\prime\prime}}}

\newcommand{\bfn}{{\bf n}}
\newcommand{\bfr}{{\bf r}}
\newcommand{\bfrp}{{\bf r}^{\prime}}
\newcommand{\bfrpp}{{\bf r}^{\prime\prime}}
\newcommand{\bfx}{{\bf x}}
\newcommand{\bfy}{{\bf y}}

\newcommand{\bfT}{{\bf T}}

\newcommand{\balpha}{ {\bm\alpha} }

\newcommand{\bmu}{\bm\mu}
\newcommand{\thrj}[6]{
        \left(
        \begin{array}{ccc}
        #1  & #2  & #3   \\
        #4  & #5  & #6   \\
        \end{array}
        \right)
        }

\begin{document}
\title{
Evaluation of the screened vacuum-polarization corrections to the hyperfine splitting of Li-like bismuth}
\author{
O.~V.~Andreev,$^{1,2}$ D.~A.~Glazov,$^{1,2}$ A.~V.~Volotka,$^{1,2}$
V.~M.~Shabaev,$^{1}$ and G.~Plunien$^{2}$ }
\affiliation{
$^1$
Department of Physics, St. Petersburg State University,
Oulianovskaya 1, Petrodvorets, St. Petersburg 198504, Russia \\
$^2$
Institut f\"ur Theoretische Physik, Technische Universit\"at Dresden,
Mommsenstra{\ss}e 13, D-01062 Dresden, Germany \\
}
\begin{abstract}
The rigorous calculation of the vacuum-polarization screening corrections to the hyperfine splitting
in Li-like bismuth is presented. The two-electron diagrams with electric and magnetic vacuum-polarization
loops are evaluated to all orders in $\alpha Z$, including the Wichmann-Kroll contributions.
This improves the accuracy of the theoretical prediction for the specific difference
of the hyperfine splitting values of H- and Li-like bismuth.
\end{abstract}

% \pacs{12.20.Ds, 31.30.Jv, 31.30.Gs}

\maketitle

%%%%%%%%%%%%%%%%%%%%%%%%%%%%%%%%%%%%%%%%%%%%%%%%%%%%%%%%%%%%%%%%%%%%%%%%%%%%%%%%%%%%%%%%%%%%%%%%%%%%
%
\section{Introduction}

High-precision measurements of the ground-state hyperfine splitting (HFS) were performed for various heavy H-like ions,
including $^{209}$Bi, $^{165}$Ho, $^{185}$Re, $^{207}$Pb, $^{203}$Tl and $^{205}$Tl
\cite{kl-prl-73,cr-prl-77,cr-pra-57,se-prl-81,be-pra-64}. Progress in experiments motivated
intensive theoretical calculations of the hyperfine splitting in highly charged heavy ions
\cite{sc-pra-50,sh-jpb-27,sh-pra-52,persson:96:prl,bl-pra-55,sh-pra-56,su-pra-58,shabaev:98:hi,boucard:00:epjd,
shabaev:00:hi,sa-pra-63,sh-prl-86,ar-pra-63,ye-pra-64} aiming to test quantum electrodynamics (QED) in strong electromagnetic fields.
It was found that in heavy ions the QED effects are obscured by the uncertainty of the nuclear magnetization
distribution correction (Bohr-Weisskopf effect).
However, simultaneous study of H- and Li-like ions of the same isotope can help to overcome this problem,
since the uncertainty of the Bohr-Weisskopf effect is significantly reduced in the specific difference
of the corresponding HFS values \cite{sh-prl-86}.
High-precision measurements of the hyperfine splitting of H- and Li-like bismuth
are feasible at the experimental storage ring (ESR) and the HITRAP facility in GSI \cite{an-hi-196,no-hi-199}.
Recently, after 13 years of attempts, the HFS of the ground state Li-like Bi has been directly observed in GSI \cite{no-hi-pc}.
These measurements together with accurate theoretical calculations will provide the possibility for the stringent tests of QED in strong fields.

The recent improvements of the theoretical accuracy for the specific difference of the HFS 
values are related to the calculations of the screened QED corrections \cite{vo-prl-103,gl-pra-81} and the two-photon exchange corrections \cite{vo-prl-2011} to the HFS of the Li-like ions. Now the uncertainty of the specific HFS difference is mainly determined by the Wichmann-Kroll part of the screened QED corrections.
In Refs. \cite{sa-pra-63,sapirstein:08:pra,or-oas-102,ko-pra-76,or-pla-372,vo-pra-78} the screened QED corrections were evaluated by introducing an effective local screening potential in the zeroth-order (Dirac) equation.
However, the screening potential approximation does not provide reliable estimation of the uncertainty.
Recently, the two-electron self-energy diagrams and a dominant part of the two-electron
vacuum-polarization diagrams, which represent the leading contribution to this effect,
have been evaluated within the systematic QED approach \cite{vo-prl-103,gl-pra-81}.
The present paper is devoted to the rigorous evaluation of the two-electron vacuum-polarization diagrams,
which have been treated approximately in Refs. \cite{vo-prl-103,gl-pra-81}.
In particular, the electric-loop and magnetic-loop diagrams are evaluated to all orders in $\alpha Z$,
including the Wichmann-Kroll contributions. The Wichmann-Kroll terms of the remaining internal-loop contributions are estimated.
The numerical results are presented for the hyperfine structure of Li-like bismuth $^{209}\textrm{Bi}^{80+}$.

The relativistic units ($\hbar=1$, $c=1$, $m=1$) and the Heaviside charge unit \nobreak[${\alpha=e^2/(4\pi),e<0}$]
are used throughout the paper.

%%%%%%%%%%%%%%%%%%%%%%%%%%%%%%%%%%%%%%%%%%%%%%%%%%%%%%%%%%%%%%%%%%%%%%%%%%%%%%%%%%%%%%%%%%%%%%%%%%%%
%
\section{Basic formulas}
\label{sec:basic}
The interaction of atomic electrons with the nuclear magnetic moment is described by the Fermi-Breit operator,
\begin{equation}
  H_\mu = \frac{|e|}{4\pi} \bmu \cdot \bfT
\,,
\end{equation}
where $\bmu$ is the operator of the nuclear magnetic moment. The electronic operator $\bfT$ is given by
\begin{equation}
  \bfT = \sum_i \frac{[\bfn_i \times \balpha_i]}{r_i^2} F(r_i)
\,,
\end{equation}
where the summation runs over the atomic electrons, $\balpha$ is the Dirac-matrix vector, $\bfn_i = \bfr_i/r_i$,
and $F(r)$ is the nuclear magnetization distribution factor (see, e.g., Refs. \cite{sh-pra-56,vo-pra-78}).
Here we employ the homogeneous sphere model,
\begin{equation}
\label{fnmd}
  F(r) =
    \begin{cases}
      (r/R_0)^3, & \mbox{if } r \leq R_0 \,,\\
      1, & \mbox{if } r > R_0 \,,
    \end{cases}
\end{equation}
where $R_0$ is the radius of the sphere, related to the root-mean-square charge radius ${\langle r^2\rangle}^{1/2}$
of the nucleus as $R_0=\sqrt{5/3}{\langle r^2 \rangle}^{1/2}$.
The ground-state hyperfine splitting of a highly charged Li-like ion in the non-recoil limit can be written as
\begin{eqnarray}
\Delta E_{\text{hfs}}^{(a)} = \frac{\alpha (\alpha Z)^3}{12} \frac{g_I}{m_p} (2I+1)
&& \left[ A(\alpha Z)(1-\delta)(1-\varepsilon) + \frac{1}{Z}B(\alpha Z) + \frac{1}{Z^2}C(\alpha Z)
\right.
\nonumber\\ 
&& \left. + \frac{1}{Z^3}D(Z, \alpha Z) + x_\text{QED} + x_\text{SQED} \right]\,.
\end{eqnarray}
Here $g_I=\mu/(\mu_N I)$ is the $g$ factor of the nucleus with magnetic moment $\mu$ and spin $I$,
$\mu_N$ is the nuclear magneton, and $m_p$ denotes the proton mass.
$A(\alpha Z)$ is the one-electron relativistic factor, $\delta$ and $\varepsilon$ are the corrections
due to the finite distribution of the charge and the magnetic moment over the nucleus, respectively, which can be found either analytically \cite{sh-jpb-27,vo-epjd-23} or numerically.
The interelectronic-interaction correction of first order in $1/Z$ is represented by the function $B(\alpha Z)$.
The function $C(\alpha Z)$ incorporates the interelectronic-interaction corrections in the second order in $1/Z$, $D(Z, \alpha Z)$ corresponds to the third- and higher-order corrections in $1/Z$.
$x_\text{QED}$ and $x_\text{SQED}$ correspond to the one-electron and many-electron (screened) QED corrections, respectively.

\begin{figure}
\includegraphics{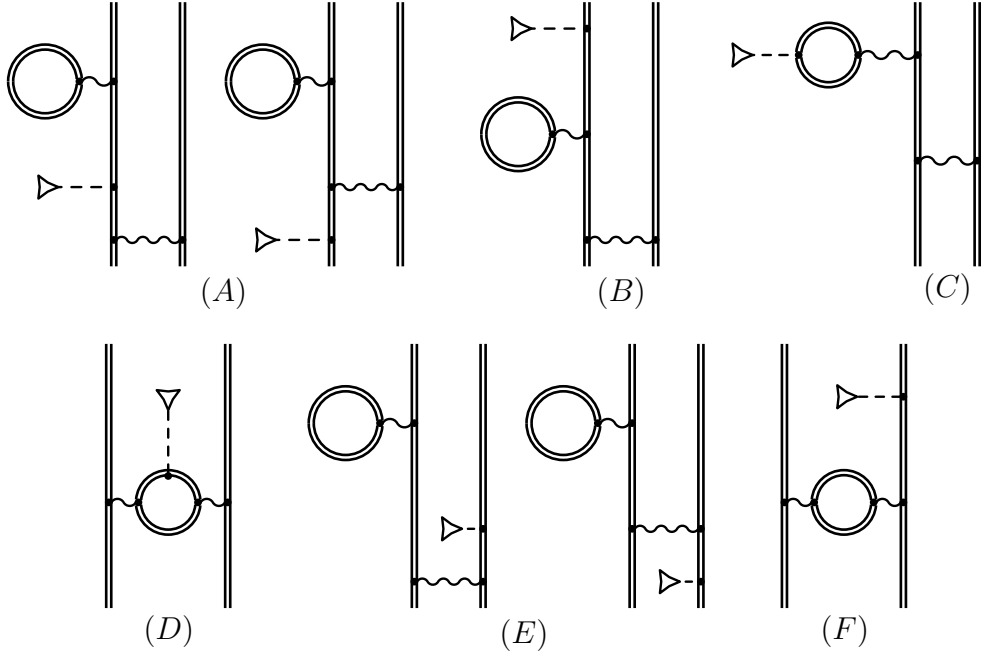}
\caption{\label{Scrvp}
Diagrams, contributing to the screened VP correction to the hyperfine splitting.
The wavy line indicates the photon propagator and the double line indicates the electron propagator in the Coulomb field.
The dashed line terminated by the triangle denotes the interaction with the nuclear magnetic field.}
\end{figure}

To first order in $\alpha$ and $1/Z$, the screened QED correction $x_\text{SQED}$ to the hyperfine splitting
is given by the sum of the self-energy (SE) and vacuum-polarization (VP) parts
\begin{equation}
  x_\text{SQED} = x_\text{SQED}^\text{SE} + x_\text{SQED}^\text{VP}.
\end{equation}
In the present paper the vacuum-polarization part $x_\text{SQED}^\text{VP}$ is considered.
The corresponding diagrams are depicted in Fig.~\ref{Scrvp}.
The total contribution of these diagrams is conveniently divided into reducible and irreducible parts.
The irreducible parts of each diagram A--F are denoted by the same letter: $x_\text{SQED}^\text{VP(A-F)}$.
The reducible contributions are to be considered together with the non-diagram terms (see Ref. \cite{gl-pra-81} for details).
They are divided into three parts, G, H, and I, according to the type of the vacuum-polarization loop:
G is associated with the electric-loop diagrams A, B, and E; H corresponds to the magnetic-loop diagram C;
and I --- to the internal-loop diagram F.

The total correction due to the two-electron vacuum-polarization is given by the sum
\begin{eqnarray}
\label{eq:xsvp}
  x_\text{SQED}^\text{VP}
    &=& x_\text{SQED}^\text{VP(A)} + x_\text{SQED}^\text{VP(B)} + x_\text{SQED}^\text{VP(C)}
      + x_\text{SQED}^\text{VP(D)} + x_\text{SQED}^\text{VP(E)} + x_\text{SQED}^\text{VP(F)}
\nonumber\\
   && + x_\text{SQED}^\text{VP(G)} + x_\text{SQED}^\text{VP(H)} + x_\text{SQED}^\text{VP(I)}
\,.
\end{eqnarray}
%
%================================================================================
%
\subsection{Electric-loop diagrams}
\label{sec:el}
The terms A, B, E, and G in Eq. (\ref{eq:xsvp}) correspond to the so-called electric-loop (el) diagrams A, B and E.
The contributions of these terms are given by the following expressions \cite{vo-prl-103,gl-pra-81}
\begin{eqnarray}
  x_\text{SQED}^\text{VP(A)} = 2\, G_a \sum_b \sum_{PQ} (-1)^{P+Q}
&&  \left[
      {\sum_{n_1,n_2}}' \langle Pa| U_\text{VP}^\text{el}|n_1 \rangle
      \frac{\langle n_1| T_0|n_2 \rangle \langle n_2 Pb| I(\Delta)| Qa Qb \rangle}
           {(\varepsilon_{Pa}-\varepsilon_{n_1})(\varepsilon_{Pa}-\varepsilon_{n_2})}
\right.
\nonumber\\
&&\left.
    + {\sum_{n_1,n_2}}' \langle Pa| U_\text{VP}^\text{el}|n_1 \rangle
      \frac{\langle n_1 Pb| I(\Delta) |n_2 Qb \rangle \langle n_2 | T_0| Qa \rangle}
           {(\varepsilon_{Pa}-\varepsilon_{n_1})(\varepsilon_{Qa}-\varepsilon_{n_2})}
\right]
\,,
\label{eq:vp-a}
\end{eqnarray}
\begin{eqnarray}
  x_\text{SQED}^\text{VP(B)} = 2\, G_a \sum_b \sum_{PQ} (-1)^{P+Q}
    {\sum_{n_1,n_2}}' \frac{\langle Pa| T_0 |n_1 \rangle }{\varepsilon_{Pa}-\varepsilon_{n_1}}
      \langle n_1| U_\text{VP}^\text{el} |n_2 \rangle \frac{\langle n_2 Pb| I(\Delta)| Qa Qb \rangle}{\varepsilon_{Pa}-\varepsilon_{n_2}}
\,,
\label{eq:vp-b}
\end{eqnarray}
\begin{eqnarray}
  x_\text{SQED}^\text{VP(E)} = 2\, G_a \sum_b \sum_{PQ} (-1)^{P+Q}
&&  \left[ {\sum_{n_1,n_2}}' \langle Pa| U_\text{VP}^\text{el}|n_1 \rangle
      \frac{\langle Pb| T_0|n_2 \rangle \langle n_1 n2| I(\Delta)| Qa Qb \rangle}
           {(\varepsilon_{Pa}-\varepsilon_{n_1})(\varepsilon_{Pb}-\varepsilon_{n_2})}
\right.
\nonumber\\
&&\left.
    + {\sum_{n_1,n_2}}' \langle Pa| U_\text{VP}^\text{el}|n_1 \rangle
      \frac{\langle n_1 Pb| I(\Delta) |Qa n_2 \rangle \langle n_2 | T_0| Qb \rangle}
           {(\varepsilon_{Pa}-\varepsilon_{n_1})(\varepsilon_{Qb}-\varepsilon_{n_2})} \right]
\,,
\label{eq:vp-e}
\end{eqnarray}
\begin{eqnarray}
  x_\text{SQED}^\text{VP(G)} &=& - 2\, G_a \sum_b \sum_{PQ} (-1)^{P+Q}
    {\sum_{n_1}}' \frac{1}{(\varepsilon_{Pa}-\varepsilon_{n_1})^2}
\nonumber\\
&&    \left\{ \langle Pa| U_\text{VP}^\text{el}|n_1 \rangle \langle n_1| T_0|Pa \rangle \langle Pa Pb| I(\Delta)| Qa Qb \rangle
\right.
\nonumber\\
&&    + \langle Pa| U_\text{VP}^\text{el}|Pa \rangle \langle Pa| T_0|n_1 \rangle \langle n_1 Pb| I(\Delta)| Qa Qb \rangle
\nonumber\\
&&\left.
      + \langle Pa| T_0|Pa \rangle \langle Pa| U_\text{VP}^\text{el}|n_1 \rangle \langle n_1 Pb| I(\Delta)| Qa Qb \rangle \right\}
\nonumber\\
&&  + 2\, G_a \sum_b \sum_{PQ} (-1)^{P+Q}
      \left\{ {\sum_{n_1}}' \frac{1}{\varepsilon_{Pa}-\varepsilon_{n_1}}
\right.
\nonumber\\
&&  \times \left[ \langle Pa| U_\text{VP}^\text{el}|n_1 \rangle \langle n_1| T_0|Pa \rangle \langle Pa Pb| I'(\Delta)| Qa Qb \rangle
\right.
\nonumber\\
&&\left.
      + \langle Pa| U_\text{VP}^\text{el}|n_1 \rangle \langle n_1 Pb| I'(\Delta)| Qa Qb \rangle
        \left( \langle Qb| T_0|Qb \rangle - \langle Pb| T_0|Pb \rangle \right) \right]
\nonumber\\
&&    + \langle Pa| U_\text{VP}^\text{el}|Pa \rangle
      \left[ {\sum_{n_1}}' \frac{\langle Pa Pb| I'(\Delta)| n_1 Qb \rangle \langle n_1| T_0|Qa \rangle }
                                {\varepsilon_{Qa}-\varepsilon_{n_1}}
\right.
\nonumber\\
&&\left.\left.
      + {\sum_{n_1}}^\prime \frac{\langle Pa Pb| I'(\Delta)| Qa n_1 \rangle \langle n_1| T_0|Qb \rangle }
                                 {\varepsilon_{Qb}-\varepsilon_{n_1}} \right] \right\}
\nonumber\\
&&    +  G_a \sum_b \sum_{PQ} (-1)^{P+Q} \langle Pa| U_\text{VP}^\text{el}|Pa \rangle \langle Pa Pb| I''(\Delta)| Qa Qb \rangle
\nonumber\\
&&    \times \left( \langle Qb| T_0|Qb \rangle - \langle Pb| T_0|Pb \rangle \right)
\,.
\label{eq:vp-g}
\end{eqnarray}
Here and below (Eqs. (\ref{eq:vp-c}), (\ref{eq:vp-h}), (\ref{eq:vp-f}), (\ref{eq:vp-i}))
$P$ and $Q$ are the permutation operators, interchanging $a$ and $b$, $\Delta \equiv \varepsilon_{Qb}-\varepsilon_{Qa}$.
The summation over $b$ runs over two core electron states with different projections of the angular momentum.
The prime at the summation sign ${\sum}'$ indicates that the energy of the intermediate state differs from the energy of the initial state,
so that the corresponding denominator is non-zero.
The interelectronic-interaction operator $I(\Delta)$ and its derivatives are defined as in Ref. \cite{sh-pr-356}.
The factor $G_a$ is defined by the quantum numbers of the valence state,
\begin{equation}
  G_a = \frac{n^3 (2l+1) j (j+1)}{2 (\alpha Z)^3 m_j} = \frac{3}{(\alpha Z)^3 m_j}
\,,
\end{equation}
where $m_j$ is the projection of the angular momentum $j$.

\begin{figure}
\includegraphics{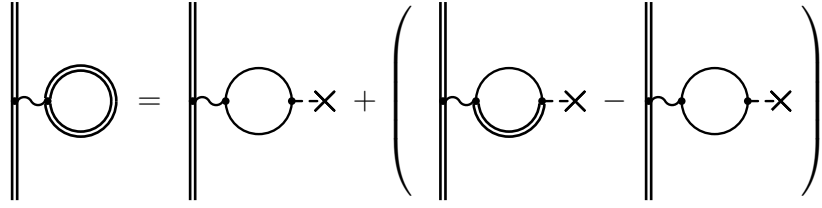}
\caption{\label{vpelexp}
The decomposition of the $U_{\text{VP}}^{\text{el}}$ into the Uehling (Ue) and
the Wichmann-Kroll (WK) terms.}
\end{figure}

Equations
(\ref{eq:vp-a})--(\ref{eq:vp-g})
involve the matrix elements of the standard electric-field-induced vacuum-polarization potential $U_\text{VP}^{\text{el}}$.
The unrenormalized expression for $U_\text{VP}^{\text{el}}$ is given by
\begin{equation}
  U_{\text{VP}}^{\text{el}}(\bfr) = \frac{\alpha}{2\pi i} \int d^3\bfrp \,\frac{1}{|\bfr-\bfrp|}
    \int_{-\infty}^\infty d\omega \,\text{Tr} [ G(\omega, \bfrp, \bfrp) ]
\,,
\label{eq:unren_vp}
\end{equation}
where $G(\omega,\bfr,\bfrp)$ is the Dirac-Coulomb Green function
\begin{equation}
  G(\omega,\bfr,\bfrp) = \sum_n \frac{\psi_n(\bfr) \psi_n^\dagger(\bfrp)}{\omega-\varepsilon_n(1-i0)}
\,.
\end{equation}
The decomposition of the vacuum-polarization loop into the Uehling (Ue) and
the Wichmann-Kroll (WK) terms is depicted in Fig.~\ref{vpelexp}.
In this expansion only the lowest order term, the Uehling term, is divergent.
The charge renormalization yields a finite well known renormalized expression
\begin{eqnarray}
  U_{\text{VP}}^{\text{el-Ue}}(r) &=&
    - \alpha Z\, \frac{2\alpha}{3\pi} \int_0^\infty d\rp \, 4\pi\rp \, \rho(\rp)
    \int_1^\infty dt \, \left(1+\frac{1}{2t^2}\right) \frac{\sqrt{t^2-1}}{t^2}
\nonumber\\
   && \times \frac{1}{4rt} \left[\exp(-2|r-\rp|t)-\exp(-2(r+\rp)t)\right]
\,,
\label{eq:el-ue}
\end{eqnarray}
where the density of the nuclear charge distribution $\rho(r)$ is normalized to 1.
The Wichmann-Kroll part (the term in brackets in the decomposition in Fig.~\ref{vpelexp})
is calculated to all orders in $\alpha Z$ as the difference between the unrenormalized total and Uehling contributions.
The resulting expression is free from divergencies, completely isolated in the Uehling term \cite{mo-pr-293}.
It is known, that no spurious terms contribute, if the calculation is based on the the partial-wave expansion
of the electron Green function and the summation is terminated after a finite number of terms \cite{gy-npa-244,ri-pra-12,so-pra-38}.
Thus the WK contribution after Wick-rotation of the contour of $\omega$-integration in the complex plane
can be written as \cite{ar-pra-60}
\begin{eqnarray}
  U_{\text{VP}}^{\text{el-WK}}(r) &=&
    \frac{2\alpha}{\pi} \sum_{\kappa} |\kappa| \int_0^\infty d\omega
    \int_0^\infty d\rp \, \rp^2 \int_0^\infty d\rpp \, \rpp^2 \frac{1}{\text{max}(r,\rp)} V(\rpp)
\nonumber\\
  && \times \sum_{i,k = 1}^2 \text{Re} \left\{ F_\kappa^{ik} (i\omega,\rp,\rpp)
    [ G_\kappa^{ik} (i\omega,\rp,\rpp) - F_\kappa^{ik} (i\omega,\rp,\rpp) ] \right\}
\,.
\end{eqnarray}
Here $G_\kappa^{ik}$ and $F_\kappa^{ik}$ are the radial components of the partial-wave contributions
to the bound and free electron Green's functions, respectively, and $V(r)$ is the electric potential of the extended nucleus.
%
%================================================================================
%
\subsection{Magnetic-loop diagrams}
\label{sec:ml}
\begin{figure}
\includegraphics{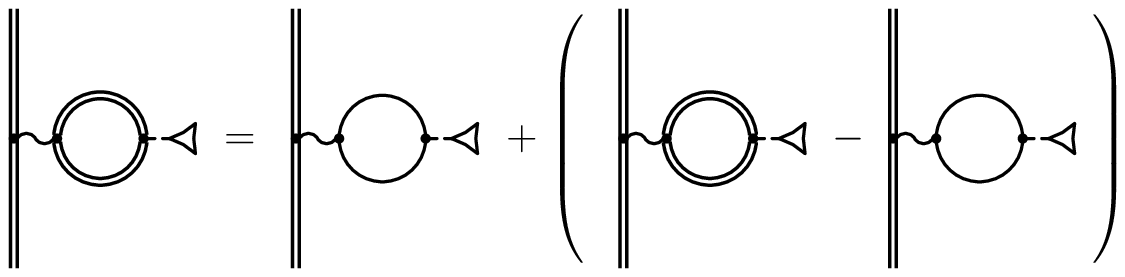}
\caption{\label{vpmlexp}
The decomposition of the $U_{\text{VP}}^{\text{ml}}$ into the Uehling and the Wichmann-Kroll terms.}
\end{figure}

The magnetic-loop (ml) diagram C corresponds to the terms C and H in Eq.~(\ref{eq:xsvp}), given by \cite{vo-prl-103,gl-pra-81}
\begin{eqnarray}
  x_\text{SQED}^\text{VP(C)} = 2\, G_a \sum_b \sum_{PQ} (-1)^{P+Q} {\sum_{n_1}}'
    \langle Pa| U_\text{VP}^\text{ml}|n_1 \rangle
    \frac{\langle n_1 Pb| I(\Delta) |Qa Qb \rangle }
         {\varepsilon_{Pa}-\varepsilon_{n_1}}
\,,
\label{eq:vp-c}
\end{eqnarray}
\begin{eqnarray}
  x_\text{SQED}^\text{VP(H)} = \frac{1}{2}\, G_a \sum_b \sum_{PQ} (-1)^{P+Q} \langle Pa Pb| I'(\Delta) |Qa Qb \rangle
    \left( \langle Qb| U_\text{VP}^\text{ml}|Qb \rangle - \langle Pb| U_\text{VP}^\text{ml}|Pb \rangle \right)
\,.
\label{eq:vp-h}
\end{eqnarray}
Equations (\ref{eq:vp-c}), (\ref{eq:vp-h})
involve the matrix elements of the magnetic-field-induced vacuum-polarization potential $U_\text{VP}^\text{ml}$.
Its unrenormalized expression reads
\begin{eqnarray}
  U_{\text{VP}}^{\text{ml}} (\bfr) = \frac{\alpha}{2\pi i} \int_{-\infty}^\infty d\omega
    \int d^3\bfrp \int d^3\bfrpp \frac{\balpha}{|\bfr-\bfrp|}
    \text{Tr} [ \balpha G(\omega, \bfrp, \bfrpp) T_0 (\bfrpp) G(\omega, \bfrpp, \bfrp) ]
\,.
\label{eq:unren_vp-ml}
\end{eqnarray}
The scalar product is implicit in Eq.~(\ref{eq:unren_vp-ml}).
Similar to the electric-loop potential, the decomposition of the magnetic-loop potential into the first order term and the remainder
(Fig. \ref{vpmlexp}) leads to the isolation of the divergency in the leading Uehling term.
It is given by the Eq.~(\ref{eq:unren_vp-ml}) with the bound-electron Green function replaced by the free-electron one.
For the sphere model of the nuclear magnetization distribution ($F(r)$ given by Eq.~(\ref{fnmd}))
the analytical expression for the renormalized magnetic-loop Uehling term reads \cite{vo-pra-78}
\begin{eqnarray}
  U_{\text{VP}}^{\text{ml-Ue}}(\bfr) = \frac{\alpha}{\pi} \frac{[\textbf n \times \balpha]_0}{r^2} \frac{3}{16R_0^3} \,
&&  \left[ 4 r R_0 [\beta_1(R_0+r) + \beta_1(|R_0-r|)]
\right.
\nonumber\\
&&  + 2(R_0+r)\beta_2(R_0+r) - 2|R_0-r|\beta_2(|R_0-r|)
\nonumber\\
&&\left.
    + \beta_3(R_0+r)-\beta_3(|R_0-r|) \right]
\,,
\label{eq:ml-ue}
\end{eqnarray}
where the function $\beta_n$ is defined by
\begin{equation}
  \beta_n(r)=\frac{2}{3} \int_1^\infty dt \,\frac{\sqrt{t^2-1}}{t^{n+2}} \left(1+\frac{1}{2t^2}\right) \exp(-2tr)
\,.
\end{equation}
Similar to the case of electric-loop term, the magnetic-loop Wichmann-Kroll contribution is calculated by summing up
the partial-wave differences between the unrenormalized total and Uehling contributions.
The magnetic-loop diagram contributes also to the nuclear magnetic moment.
The corresponding Uehling term is equal to zero, however the Wichmann-Kroll term is not.
Therefore, in the calculations of the WK contribution one should account for the related contribution
to the nuclear magnetic moment in the zeroth-order HFS value (see Refs. \cite{su-pra-58,ar-pra-63,mi-plb-233}).
This implies the replacement of the nuclear magnetic moment $\mu$ by the `bare' value
\begin{equation}
  \mu \rightarrow \mu_\text{bare}=\mu - \Delta \mu
\,.
\end{equation}
The nuclear magnetic moment correction $\Delta \mu$ due to the magnetic-loop WK part
can be expressed as $\Delta \mu = \epsilon \mu$ with the dimensionless parameter $\epsilon$ given by
\begin{eqnarray}
  \epsilon = \frac{1}{2\pi i} \frac{\alpha}{2} \int d^3\bfr \int d^3\bfrp \int_{-\infty}^\infty d\omega
&&  \left[ \text{Tr} \left\{ [\bfr \times \balpha]_0 G(\omega,\bfr,\bfrp)
      T_0(\bfrp) G(\omega,\bfrp,\bfr) \right\}
\right.
\nonumber\\
&&
\left.
         - \text{Tr} \left\{ [\bfr \times \balpha]_0 F(\omega,\bfr,\bfrp)
      T_0(\bfrp) F(\omega,\bfrp,\bfr) \right\}
\right]
\,.
\end{eqnarray}
Finally, the corrected magnetic-loop WK contribution is obtained by subtraction of $\epsilon T_0$
from the WK part of Eq.~(\ref{eq:unren_vp-ml}). The corresponding expression reads
\begin{eqnarray}
  U_{\text{VP}}^{\text{ml-WK}} (\bfr) &=& \frac{\alpha}{2\pi i} \int_{-\infty}^\infty d\omega \int d^3\bfrp \int d^3\bfrpp
\nonumber\\
&& \times \left( \frac{\balpha}{|\bfr-\bfrp|} \left[
     \text{Tr} \left\{\balpha G(\omega,\bfrp,\bfrpp) T_0 (\bfrpp) G(\omega,\bfrpp,\bfrp)\right\}
\right.\right.
\nonumber\\
&&\left.
     - \text{Tr} \left\{\balpha F(\omega,\bfrp,\bfrpp) T_0 (\bfrpp) F(\omega,\bfrpp,\bfrp)\right\} \right]
\nonumber\\
&&
     - \frac{1}{2}\, T_0(\bfr) \left[
       \text{Tr} \left\{[\bfrp \times \balpha]_0 G(\omega,\bfrp,\bfrpp) T_0 (\bfrpp) G(\omega,\bfrpp,\bfrp)\right\}
\right.
\nonumber\\
&&\left.\left.
     - \text{Tr} \left\{[\bfrp \times \balpha]_0 F(\omega,\bfrp,\bfrpp) T_0 (\bfrpp) F(\omega,\bfrpp,\bfrp)\right\} \right]
\vphantom{\frac{\balpha}{|\bfx-\bfy|}}
\right)
\,.
\end{eqnarray}
The angular-momentum conservation allows one to distinguish in this expression
the following angular dependence
\begin{equation}
  U_{\text{VP}}^{\text{ml-WK}} (\bfr) =  [\bfn \times \balpha]_0 \, u_{\text{VP}}^{\text{ml-WK}} ( r )
\,.
\end{equation}
The radial part $u_{\text{VP}}^{\text{ml-WK}}(r)$ can be presented in the form
\begin{equation}
  u_{\text{VP}}^{\text{ml-WK}} (r) = \int_0^\infty d\rp \rp^2 \left(\frac{r_<}{r_>^2} - \frac{\rp}{r^2}\, F(r) \right) \rho_\text{ml}(\rp)
\,,
\end{equation}
where the magnetic-loop WK charge density $\rho_\text{ml}(\rp)$ after rotating the contour of $\omega$-integration
can be written as
\begin{eqnarray}
  \rho_\text{ml}(\rp) = \frac{\alpha}{6\pi} \int_0^\infty d\rpp F(\rpp) \int_0^\infty d\omega \sum_{\kappa_1,\kappa_2}
    B^2 (\kappa_1,\kappa_2) S_{\kappa_1 \kappa_2} ( \omega, \rp, \rpp )
\,.
\label{eq:rho-mag}
\end{eqnarray}
Here the Green-function trace $S_{\kappa_1 \kappa_2}$ is given by
\begin{eqnarray}
  S_{\kappa_1 \kappa_2} ( \omega, \rp, \rpp ) = &\text{Re} &\Big\{
      G^{11}_{\kappa_1}(i\omega, \rp, \rpp) \, G^{22}_{\kappa_2}(i\omega, \rp, \rpp)
    + G^{12}_{\kappa_1}(i\omega, \rp, \rpp) \, G^{21}_{\kappa_2}(i\omega, \rp, \rpp)
\nonumber\\
&&  + G^{21}_{\kappa_1}(i\omega, \rp, \rpp) \, G^{12}_{\kappa_2}(i\omega, \rp, \rpp)
    + G^{22}_{\kappa_1}(i\omega, \rp, \rpp) \, G^{11}_{\kappa_2}(i\omega, \rp, \rpp)
\nonumber\\
&&  - F^{11}_{\kappa_1}(i\omega, \rp, \rpp) \ F^{22}_{\kappa_2}(i\omega, \rp, \rpp)
    - F^{12}_{\kappa_1}(i\omega, \rp, \rpp) \ F^{21}_{\kappa_2}(i\omega, \rp, \rpp)
\nonumber\\
&&  - F^{21}_{\kappa_1}(i\omega, \rp, \rpp) \ F^{12}_{\kappa_2}(i\omega, \rp, \rpp)
    - F^{22}_{\kappa_1}(i\omega, \rp, \rpp) \ F^{11}_{\kappa_2}(i\omega, \rp, \rpp) \Big\}
\,,
\end{eqnarray}
and the angular factor $B(\kappa_1,\kappa_2)$ is defined by the following expression
\begin{equation}
\label{hfs-angle}
  B(\kappa_1,\kappa_2)=\frac{1+(-1)^{l_1+l_2}}{2}\sqrt{2(2j_1+1)(2j_2+1)}\,(-1)^{l_1}
    \thrj{j_1}{j_2}{1}{\frac{1}{2}}{\frac{1}{2}}{-1}
\,.
\end{equation}
The selection rule, following from this expression, forces $\kappa_2$ to be equal to either $\kappa_1$ or $-\kappa_1 \pm 1$.
Therefore the double $\kappa$-summation can be reduced to a single one, including diagonal and off-diagonal $\kappa$-terms,
\begin{equation}
  \sum_{\kappa_1,\kappa_2} B^2 (\kappa_1,\kappa_2) S_{\kappa_1 \kappa_2} =
    \sum_{\kappa} \left[ \frac{2|\kappa|^3}{\kappa^2-1/4} S_{\kappa \kappa}
      + 2\frac{|\kappa| (|\kappa|+1)}{|\kappa|+1/2} S_{\kappa \bar\kappa} \right]
\,,
\end{equation}
where $\bar{\kappa}=-\kappa - \text{sign}(\kappa)$.
%
%===================================================================================
%
\subsection{Internal-loop diagrams}
\label{sec:il}
\begin{figure}
\includegraphics{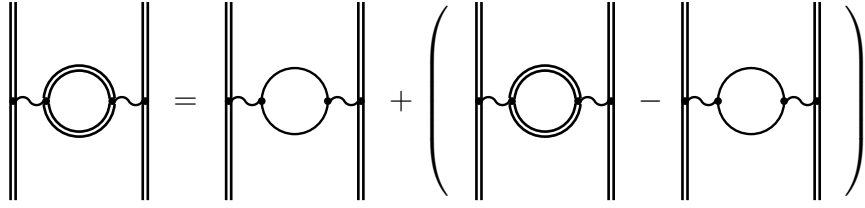}
\caption{\label{vpilexp}
The decomposition of the internal electric-loop $I_{\text{VP}}$ into the Uehling and the Wichmann-Kroll terms.}
\end{figure}

The internal-loop diagram F corresponds to the terms F and I in Eq.~(\ref{eq:xsvp}), given by \cite{vo-prl-103,gl-pra-81}
\begin{equation}
\label{eq:vp-f}
  x_\text{SQED}^\text{VP(F)} = 2 \, G_a \sum_b \sum_{PQ} (-1)^{P+Q} {\sum_{n_1}}^\prime
    \frac{\langle Pa| T_0 |n_1 \rangle}{\varepsilon_{Pa}-\varepsilon_{n_1}} \langle n_1 Pb| I_\text{VP}(\Delta) |Qa Qb \rangle
\,,
\end{equation}
\begin{eqnarray}
\label{eq:vp-i}
  x_\text{SQED}^\text{VP(I)} = \frac{1}{2} \, G_a \sum_b \sum_{PQ} (-1)^{P+Q}
    \langle Pa Pb| I_\text{VP}'(\Delta) |Qa Qb \rangle \left( \langle Qb| T_0 |Qb \rangle - \langle Pb| T_0 |Pb \rangle \right)
\,.
\end{eqnarray}
Equations (\ref{eq:vp-f}) and (\ref{eq:vp-i})
involve the interelectronic-interaction operator, modified by the vacuum-polarization loop,
\begin{eqnarray}
  I_{\text{VP}}(\varepsilon,\bfr_1,\bfr_2) &=& \frac{\alpha^2}{2\pi i} \int_{-\infty}^\infty d\omega \int d^3\bfrp_1 \int d^3\bfrp_2 \,
    \frac{\alpha_{1\mu} \exp (i|\varepsilon||\bfr_1-\bfrp_1|)}{|\bfr_1-\bfrp_1|} \,
    \frac{\alpha_{2\nu} \exp (i|\varepsilon||\bfr_2-\bfrp_2|)}{|\bfr_2-\bfrp_2|}
\nonumber\\
&&  \times \text{Tr} \left[ \alpha^\mu G(\omega-\varepsilon/2, \bfrp_1, \bfrp_2) \alpha^\nu G(\omega+\varepsilon/2, \bfrp_2, \bfrp_1) \right]
\,,
\label{eq:unren_vp-ii}
\end{eqnarray}
where $\varepsilon$ is the energy of the transmitted photon, $\alpha^{\mu}=(1,\balpha)$ is the four-vector of the Dirac matrices,
and the summation over $\mu$ and $\nu$ is implicit.
The matrices $\alpha_{1\mu}$ and $\alpha_{2\nu}$ act on the spinor variables corresponding to $\bfr_1$ and $\bfr_2$, respectively.
This operator is also divided into the leading divergent Uehling part and the remaining finite Wichmann-Kroll part
(see Fig.~\ref{vpilexp}).
The renormalized expression for the Uehling term reads (see, e.g., \cite{ar-pra-60})
\begin{eqnarray}
  I_\text{VP}^{\text{Ue}}(\varepsilon,\bfr_1,\bfr_2) =
     \frac{2\alpha^2}{3\pi} \, \frac{\alpha_{1\mu}\alpha^\mu_2}{|\bfr_1-\bfr_2|} \,
    \int_1^\infty dt \left(1+\frac{1}{2t^2}\right) \frac{\sqrt{t^2-1}}{t^2}
    \exp\left( - \sqrt{4t^2-\varepsilon^2} \, |\bfr_1-\bfr_2| \right)
\,.
\label{eq:ii-ue}
\end{eqnarray}
The corresponding contribution is taken into account rigorously.
The direct part of the Wichmann-Kroll contribution with the hyperfine interaction vertex on the $2s$-electron 
line has been calculated by introducing the screening potential of the $(1s)^2$ closed shell electrons  
into the electric loop.
We take the difference between the contributions
with the Green function inside the loop calculated with and without the screening potential.
The calculation of the direct part of the diagram D, which  involves the internal loop modified by the hyperfine interaction vertex,
has been performed in a similar way.
It is evaluated by including the potential of the closed shell into the Green functions of 
the internal loop, that leads to a diagram of the magnetic-loop type.
The evaluation of the remaining exchange diagrams is currently underway.
%
%===================================================================================
%
\section{Results and discussion}
\label{sec:results}

The contribution of the two-electron vacuum-polarization diagrams to the hyperfine splitting of Li-like ion is calculated
in coordinate space according to the formulas presented above.
The electric-loop terms (A, B, E, G) and the magnetic-loop terms (C and H) are taken into account completely,
including the Uehling and the Wichmann-Kroll parts.
The internal-loop diagram F and the corresponding reducible term I are calculated rigorously in the Uehling approximation.
The related WK contributions are evaluated for the direct parts with the hyperfine interaction vertex on the $2s$-electron.
The remaining terms are estimated employing the assumption that the ratio of the WK part to the corresponding Uehling
part is the same. The uncertainty is ascribed as high as $200\%$ of 
the value obtained in this way.
The contribution of the direct part of the diagram D is evaluated as described in the previous section.
We ascribe a $200\%$ uncertainty to the D contribution as a conservative estimation of the uncalculated exchange term.
The numerical evaluation of the one-electron wave functions for the initial ($a$) and intermediate ($n_{1,2}$) states is performed
using the dual kinetic balance (DKB) approach \cite{sh-prl-04} with the basis set constructed from the B-splines \cite{sapirstein:96:jpb}.
The Fermi model for the nuclear charge distribution is employed in these calculations.
The Uehling parts of the vacuum-polarization potentials are calculated according to the expressions (\ref{eq:el-ue}), (\ref{eq:ml-ue}),
and (\ref{eq:ii-ue}).
The Wichmann-Kroll parts involve the free- and bound-electron Green functions for the Dirac equation.
The spherical shell model (possessing analytical solution for the bare nucleus case) and Fermi model of the nuclear charge distribution
were employed in calculations of the Green function.
The nuclear magnetization distribution effect is taken into account within the homogeneous sphere model.
The partial-wave expansion is terminated at $|\kappa|=5$ in case of the electric-loop Wichmann-Kroll parts
and at $|\kappa|=10$ in case of the magnetic-loop Wichmann-Kroll parts.
The remainders of the k-summations are estimated using the least-square inverse-polynomial fitting.
The numerical calculation procedure has been performed in different gauges, the gauge invariance should hold for the complete set of the two-electron
vacuum-polarization diagrams. The total values obtained in the Feynman and Coulomb gauges for the photon propagator
mediating the interelectronic interaction agree within the level of the numerical accuracy.
The results for Li-like bismuth $^{209}\text{Bi}^{80+}$ in both gauges are presented term by term in Table \ref{scrvptable1}.
\begin{table}
\caption{\label{scrvptable1}
Screened VP corrections to the HFS of Li-like bismuth ${}^{209}\!\text{Bi}^{80+}$
in terms of $x_{\text{SQED}}$ for the Coulomb and Feynman gauges. Ue $=$ Uehling, WK $=$ Wichmann-Kroll.}
\begin{center}
\begin{tabular}{cr@{}lr@{}lr@{}lr@{}l}
\hline
\hline
& \multicolumn{4}{c}{Feynman} & \multicolumn{4}{c}{Coulomb} \\
& \multicolumn{2}{c}{Ue} & \multicolumn{2}{c}{WK} & \multicolumn{2}{c}{Ue} & \multicolumn{2}{c}{WK} \\
\hline
A           & $-$0&.000 5074 &    0&.000 0178 & $-$0&.000 5085 &    0&.000 0179 \\
B           & $-$0&.000 2192 &    0&.000 0056 & $-$0&.000 2166 &    0&.000 0055 \\
C           & $-$0&.000 1692 &    0&.000 0461 & $-$0&.000 1670 &    0&.000 0455 \\
D           & \multicolumn{2}{c}{$-$} & 0&.000 0021 & \multicolumn{2}{c}{$-$} & 0&.000 0021\\
E           & $-$0&.000 0033 &    0&.000 0002 & $-$0&.000 0031 &    0&.000 0002 \\
F           &    0&.000 0015 & $-$0&.000 0002 &    0&.000 0015 & $-$0&.000 0002 \\
G           &    0&.000 2896 & $-$0&.000 0123 &    0&.000 2879 & $-$0&.000 0124 \\
H           &    0&.000 0023 & $-$0&.000 0006 &    0&.000 0001 & $-$0&.000 0000 \\
I           &    0&.000 0000 &    0&.000 0000 &    0&.000 0000 &    0&.000 0000 \\
\hline
Total (A-I) & $-$0&.000 6056 &    0&.000 0586 & $-$0&.000 6056 &    0&.000 0586 \\
\hline
Total (Ue+WK) & \multicolumn{4}{c}{$-$0.000 5470} & \multicolumn{4}{c}{$-$0.000 5470} \\
\hline
\hline
\end{tabular}
\end{center}
\end{table}

\begin{table}
\caption{\label{Scrvptable2}
Screened VP corrections to the HFS of Li-like bismuth ${}^{209}\textrm{Bi}^{80+}$ in terms of $x_{\textrm{SQED}}$
in the Feynman gauge are compared to the corresponding corrections from  Ref. \cite{gl-pra-81}.
The Wichmann-Kroll part of the magnetic-loop diagram (WK-ml)
is separated for comparison with the estimation of Ref. \cite{gl-pra-81}.}
\begin{center}
\begin{tabular}{cr@{}lr@{}l}
\hline
\hline
& \multicolumn{2}{c}{This work} & \multicolumn{2}{c}{Ref. \cite{gl-pra-81}} \\
\hline
A           & $-$0&.000 4897    & $-$0&.000 4881 \\
B           & $-$0&.000 2136    & $-$0&.000 2128 \\
C           & $-$0&.000 1692    & $-$0&.000 1691 \\
D           &    0&.000 0021(42)&     &          \\
E           & $-$0&.000 0031    & $-$0&.000 0031 \\
F           &    0&.000 0013(3) &    0&.000 0015 \\
G           &    0&.000 2773    &    0&.000 2766 \\
H           &    0&.000 0023    &    0&.000 0023 \\
I           &    0&.000 0000    &    0&.000 0000 \\
WK-ml       &    0&.000 0454    &    0&.000 05(2) \\
\hline
Total       & $-$0&.000 547(4)  & $-$0&.000 54(2) \\
\hline
\hline
\end{tabular}
\end{center}
\end{table}
In Table \ref{Scrvptable2} the individual terms are compared with those from Ref. \cite{gl-pra-81}.
For the electric-loop terms the sums of the Uehling and Wichmann-Kroll contributions are presented.
In Ref. \cite{gl-pra-81} the Wichmann-Kroll parts were calculated by means of the approximate formulas
for the electric WK-potential from Ref. \cite{fa-jpb-23}. 
For the magnetic-loop terms (C and H) the Wichmann-Kroll part is figured out separately (WK-ml) in order to compare the result
with the estimation given in Ref. \cite{gl-pra-81}. In that work it was evaluated utilizing the hydrogenic 2s value from Ref. \cite{ar-pra-63},
assuming that it enters with the same screening ratio as the Uehling term.

The uncertainty of the present evaluation is determined by the uncalculated terms in sets D and F. 
It was estimated as high as $200\%$ of the calculated WK terms in these sets.

\begin{table}
\caption{\label{scrvptable3}
Individual contributions to the specific difference $\Delta^\prime E$
of the hyperfine splittings of Li-like and H-like bismuth ${}^{209}\textrm{Bi}$.
Units are meV.}
\begin{center}
\begin{tabular}{lr@{}lr@{}lr@{}l}
\hline
\hline
& \multicolumn{2}{c}{$\Delta E^{(2s)}$} & \multicolumn{2}{c}{$\xi \Delta E^{(1s)}$} & \multicolumn{2}{c}{$\Delta^\prime E$} \\
\hline
Dirac value                                                &    844&.829    & 876&.638 & $-$31&.809 \\
Interelectronic interaction, $\sim 1/\text{Z}$             & $-$ 29&.995    & \multicolumn{2}{c}{}  & $-$29&.995 \\
Interelectronic interaction,  $\sim 1/\text{Z}^2$          &      0&.258    & \multicolumn{2}{c}{}  & 0&.258 \\
Interelectronic interaction,  $\sim 1/\text{Z}^3$ and h.o. &  $-$ 0&.003(3) & \multicolumn{2}{c}{}  & $-$0&.003(3) \\
QED                                                        &  $-$ 5&.052    & $-$5&.088 &     0&.036 \\
Screened SE                                                &      0&.381    &     &     &     0&.381    \\
Screened VP, this work                                     &  $-$ 0&.188(2) &     &     & $-$ 0&.188(2) \\
Screened VP, Refs. \cite{vo-prl-103,gl-pra-81}             &  $-$ 0&.187(6) &     &     & $-$ 0&.187(6) \\
Total                                                      &
\multicolumn{2}{c}{}   & \multicolumn{2}{c}{}  & $-$61&.320(4)(5) \\
\hline
\hline
\end{tabular}
\end{center}
\end{table}
In Table \ref{scrvptable3} the specific difference of the ground-state hyperfine splitting in H-like bismuth ${}^{209}\textrm{Bi}^{82+}$
and in Li-like bismuth ${}^{209}\textrm{Bi}^{80+}$, $\Delta^\prime E = \Delta E^{(2s)} - \xi \Delta E^{(1s)}$, is considered.
The parameter $\xi$ is chosen to cancel the Bohr-Weisskopf correction, $\xi=0.16886$ \cite{vo-prl-103}.
The rms radius was taken to be $\langle r^2\rangle^{1/2}=5.5211$ fm \cite{an-adndt-04}, the nuclear spin and parity $I^\pi=9/2-$,
and the magnetic moment $\mu/\mu_N=4.1106(2)$ \cite{st-adndt-05}.
The most accurate values for the interelectronic-interaction contributions are taken from the recent paper \cite{vo-prl-2011},
where the contribution of the two-photon-exchange diagrams has been evaluated in the framework of QED.
The contribution of the screened vacuum polarization calculated in this work equals $-0.188(2)$ meV
that agrees with the previous value $-0.187(6)$ meV from Refs. \cite{vo-prl-103,gl-pra-81}.
The uncertainty of the present result, being $3$ times smaller than the previous one, 
is determined by the conservative estimates
for the contributions which have not been taken into account rigorously so far. 
The first error bar in the total value of the specific difference $-61.320(4)(5)$ originates 
from the uncertainties of the screened VP contribution and the $1/\text{Z}^3$ and h.o. 
interelectronic-interaction term. The second uncertainty comes from the
nuclear magnetic moment, the nuclear polarization corrections \cite{ne-plb-552}, and 
other nuclear effects, which are not completely cancelled in the specific difference.

In summary,
calculations  of the major part of the screened vacuum-polarization correction
to the hyperfine splitting in Li-like bismuth have been performed.
The Wichmann-Kroll contributions to the two-electron electric-loop and magnetic-loop diagrams
and to the direct parts of the internal-loop diagrams
have been evaluated within the rigorous QED approach. As a result,
the accuracy of the screened QED contribution 
to the hyperfine splitting  
of Li-like bismuth has been significantly improved. 
These results, combined with the recent rigorous calculations of 
the two-photon exchange contributions \cite{vo-prl-2011}, provide a new value of
the specific difference of the HFS values in H-like and Li-like bismuth which
is by an order of magnitude more precise compared to the previous one 
 \cite{vo-prl-103,gl-pra-81}.
\acknowledgments
Valuable conversations with A.~N.~Artemyev are gratefully acknowledged.
The work reported in this paper was supported by the Deutsche Forschungsgemeinschaft (Grants No. VO 1707/1-1 and PL 254/7-1), 
by GSI, by RFBR (Grant No. 10-02-00450), by the Russian Ministry of Education and Science
(Program ``Scientific and pedagogical specialists for innovative Russia'', Grant No. P1334),
by the grant of the President of the Russian Federation (Grant No. MK-3215.2011.2),
and by the German-Russian Interdisciplinary Science Center (G-RISC) funded by the German Federal Foreign Office
via the German Academic Exchange Service (DAAD).
O.V.A. acknowledges the financial support from the ``Dynasty'' foundation and from the DAAD.
The work of D.A.G. was supported by the ``Dynasty'' foundation and by the FAIR -- Russia Research Center.
O.V.A. and D.A.G. would like to acknowledge the hospitality shown by TUD.

\end{document}